\documentstyle[aps,twocolumn,epsfig]{revtex}
\begin{document}

\newcommand{\ve}[1]{\mbox{\boldmath $#1$}}
\twocolumn[\hsize\textwidth\columnwidth\hsize
\csname@twocolumnfalse%
\endcsname
\draft
\title{Zero-temperature phase diagram of binary
boson-fermion mixtures} 

\author{L. Viverit$^{a}$, C. J. Pethick$^{b}$ 
and H. Smith$^{a}$} 

\address{ 
     {\setlength{\baselineskip}{18pt}
     $^a$ \O rsted Laboratory, H. C. \O rsted Institute, 
     Universitetsparken 5, DK-2100 Copenhagen \O, Denmark,\\
     $^b$ Nordita, Blegdamsvej 17, DK-2100 Copenhagen \O, Denmark.\\} }

\date{November 1999}
\maketitle

\begin{abstract}
\noindent 
We calculate the phase diagram for dilute
 mixtures of bosons
and  fermions  at zero temperature.
The linear stability conditions
are derived and related to the effective
boson-induced interaction between the fermions.
We show that in equilibrium there are three possibilities:
a) a single uniform phase,
b) a purely fermionic phase coexisting with a purely
bosonic one and c) a purely fermionic phase coexisting with a mixed phase.
\end{abstract}

\vskip2pc]

\section{Introduction}

Recent developments in the trapping and cooling  of atoms
have made it possible to investigate the properties of
dilute gases at very low temperatures where the
bosonic or fermionic character of the atoms
becomes crucial. Following the  realization of
 Bose-Einstein condensation in a single-component gas
of alkali atoms  \cite{bec1,bec2,bec3}, the group at JILA  succeeded in
trapping and cooling $^{87}$Rb atoms in two different hyperfine
 states, thereby creating overlapping
 condensates  in this boson-boson mixture \cite{corn}.

Lately a lot of attention has been given also to fermions, especially
in view of the possibility of achieving temperatures low enough to
observe a BCS transition.
Due to the vanishing $s$-wave cross section for identical
fermions in the same spin state, evaporative cooling of
 a single species of fermions is ineffective.
By using a mixture of two spin states DeMarco and Jin \cite{jin}
 recently succeeded in cooling
the fermionic isotope $^{40}$K to temperatures lower than the
degeneracy temperature.
Another possibility, however, is to  trap
fermions together with bosons and cool the latter,
so that the fermions are cooled through their thermal contact with the
bosons (so-called sympathetic cooling).
Such fermion-boson mixtures form the topic of the present paper.

The miscibility properties of  boson-boson  mixtures
in a trap have been discussed in several papers
 \cite{ho,ping1,ping2}. Trapped boson-fermion mixtures
have been considered in ref.\ \cite{klaus}
within  the Thomas-Fermi approximation and in ref. \cite{nicolai},
where the separation of the components was studied numerically
as a function of the interparticle interaction.
The purpose of the present work is to carry out an analytical study
of the miscibility of fermion-boson mixtures in the uniform case,
in order to obtain insight into the types of phase boundaries
that may occur in this system.

The article is  organized as follows.
We begin in section \ref{linstab} by studying linear
stability of uniform mixtures and  introduce the effective fermion-fermion
interaction as mediated by the bosons.
 In section \ref{gencom} we determine the general conditions
for phase equilibrium by considering the pressure
and chemical potentials on each side of a phase boundary.
We find that depending on the total densities of the two components, 
as shown in Fig.\ 3, 
there are three possibilities,
 a uniform phase with bosons and fermions fully mixed,
 a purely fermionic phase coexisting with a purely
bosonic one, and  a purely fermionic phase coexisting with a mixed phase.
However, a separation into two phases, each with
different, non-zero, concentrations of bosons and fermions
is never in equilibrium, nor is one
with a purely bosonic phase and a mixed one.
In section \ref{discussion} we
discuss the experimental implications
of our results, using the fermion isotopes $^6$Li, $^{40}$K and
$^{84}$Rb as examples.

\section{Linear stability and induced interactions}

\label{linstab}

The general conditions for stability of a binary mixture
towards small changes in the concentrations of its components
are obtained by considering the total energy, ${\cal E}$, as a functional
of the densities $n_{\alpha}$ and $n_{\beta}$ of the two components:
\begin{equation}
{\cal E}=\int d{\bf r}E(n_{\alpha}({\bf r}),n_{\beta}(
{\bf r})).
\end{equation}
We  consider the change in total energy
arising from small changes, $\delta n_{\alpha}$
and $\delta n_{\beta}$, in the concentrations of the two constituents.
 The first order variation, $\delta {\cal E}$, must vanish,
since the number of particles of each species is conserved,
\begin{equation}
\int d{\bf r}\delta n_i=0;\;\;\; i=\alpha,\beta.
\end{equation}
The  second order variation, $\delta^2 {\cal E}$,
is given by the quadratic form
\begin{eqnarray}
\nonumber \delta^2 {\cal E}&=&\frac{1}{2}\int d{\bf r}\left(
\frac{\partial^2 E}{\partial n_{\alpha}^2}(\delta n_{\alpha})^2+
\frac{\partial^2 E}{\partial n_{\beta}^2}(\delta n_{\beta})^2 \right.\\
&+& \left. 2\frac{\partial^2 E}{\partial n_{\alpha}\partial n_{\beta}}
\delta n_{\alpha} \delta n_{\beta}\right).
\label{quadr}
\end{eqnarray}
In carrying out this expansion we have implicitly assumed that the
characteristic wavelength for the spatial variation
of the densities is long compared to
the microscopic lengths in the problem. As we shall see, 
for dilute mixtures of bosons and fermions the relevant length
is the coherence length of the bosons.
The derivative of the energy density with respect
to the particle density, $\partial E/\partial n_i$,
is the chemical potential $\mu_i$ for the
species labelled $i$ ($i=\alpha,\beta$).
The quadratic form (\ref{quadr}) is thus positive semi-definite,
provided
\begin{equation}
\frac{\partial \mu_{\alpha}}{\partial n_{\alpha}}\geq 0,\,\,\,\,\,
\frac{\partial \mu_{\beta}}{\partial n_{\beta}}\geq 0,
\label{stab1}
\end{equation}
and
\begin{equation}
\frac{\partial \mu_{\alpha}}{\partial n_{\alpha}}\frac{\partial \mu_{\beta}}
{\partial n_{\beta}}-
\frac{\partial \mu_{\alpha}}{\partial n_{\beta}}
\frac{\partial \mu_{\beta}}{\partial n_{\alpha}}\geq 0.
\label{stab2}
\end{equation}
Note that it 
is only necessary that one of the
two conditions in (\ref{stab1}) is satisfied, since the other one is
automatically fulfilled when the condition
(\ref{stab2}) is.

For a dilute mixture of bosons with fermions in a single internal state,
at zero temperature the energy functional has the form
\begin{equation}
{\cal E}=V\left(\frac{1}{2}n_B^2U_{BB}+n_Bn_FU_{BF}+
\frac{3}{5}\epsilon_Fn_F\right),
\label{enexpre}
\end{equation}
where $V$ is the total volume. The first two terms are due to the
boson-boson and boson-fermion interactions, respectively, with $n_B$ and $n_F$
denoting the boson and fermion densities. $U_{BB}$ is the matrix element
of the effective interaction for bosons with bosons and $U_{BF}$ that
for bosons with fermions. The last term
represents the kinetic energy of the fermions. The  fermion-fermion
interaction energy is negligible,
since a) the $s$-wave scattering amplitude vanishes for
fermions in the same spin state and b) none of the higher
partial waves contribute at the low temperatures we are considering.
The chemical potentials are then seen to be
\begin{eqnarray}
\nonumber \mu_B &=& n_BU_{BB}+n_FU_{BF},\\
\mu_F &=& \epsilon_F+n_BU_{BF}.
\end{eqnarray}
Let us  introduce the constant $A$ through the definition
\begin{equation}
\epsilon_F=An_F^{2/3},\;\;\;A=\frac{\hbar^2}{2m_F}(6\pi^2)^{2/3},
\end{equation}
where $m_F$ denotes the fermion mass.
The value of the numerical factor in $A$ reflects the fact that
we consider fermions in a definite spin state. According to
(\ref{stab2}) linear stability then requires
\begin{equation}
n_F^{1/3}
\leq \frac{2}{3}\frac{AU_{BB}}{U_{BF}^2}.
\label{linsta}
\end{equation}
If the interaction parameters are converted to scattering lengths
$a_{BB}$ and $a_{BF}$ through the relations
\begin{equation}
U_{BB}=\frac{4\pi\hbar^2 a_{BB}}{m_{B}};\;\;\;
U_{BF}=\frac{4\pi\hbar^2 a_{BF}}{m_{BF}},
\end{equation}
where $m_B$ is the boson
mass and $m_{BF}=2m_Fm_B/(m_F+m_B)$
is twice the reduced mass of a boson-fermion pair,
then the condition (\ref{linsta}) becomes
\begin{equation}
n_F^{1/3}
\leq \frac{(6\pi^2)^{2/3}}{12\pi}\frac{m_{BF}^2a_{BB}}{m_Bm_Fa_{BF}^2}.
\label{linsta1}
\end{equation}
If the masses as well as the scattering lengths occurring
in (\ref{linsta1}) are approximately equal in magnitude,
then the stability condition requires  the mean interfermion distance,
which is  approximately  $n_F^{-1/3}$, to be greater than
a scattering length. The trapped one-component gases
that have so far been investigated experimentally are dilute
in the sense that the mean interparticle distance
is much greater than the scattering length.
The stability condition (\ref{linsta1}) would therefore
generally be expected to hold for dilute mixtures, unless
the scattering length  $a_{BF}$ greatly exceeds $a_{BB}$. In this
case the condition (\ref{linsta1}) may be violated at much lower
fermion densities given by $n_F^{1/3}a_{BB}\approx (a_{BB}/a_{BF})^2$
rather than $n_F^{1/3}a_{BB}\approx 1$.

It should be noted that
the mean interfermion distance in the stability condition (\ref{linsta1})
formally plays the role of an effective fermion-fermion scattering length.
This is apparent when we   consider binary mixtures of bosons,
 labelled by $\alpha$
and $\beta$. In this case the energy functional  takes the form
\begin{equation}
{\cal E}=V\left(\frac{1}{2}n_{\alpha}^2U_{\alpha\alpha}
+n_{\alpha}n_{\beta}U_{\alpha \beta}+\frac{1}{2}n_{\beta}^2U_{\beta\beta}.
\right).
\end{equation}
The condition (\ref{linsta}) would then  be replaced by
\begin{equation}
U_{\alpha\alpha}U_{\beta\beta}\geq U_{\alpha\beta}^2.
\label{bosecond}
\end{equation}
The  conditions (\ref{linsta}) and (\ref{bosecond}) assume the same
form, apart from numerical factors, provided
the combination $4\pi\hbar^2 n_F^{-1/3}/m_{F}$
is interpreted as arising from a direct fermion-fermion interaction,
while physically, of course, the term originates in the kinetic energy.
The Fermi pressure thus
behaves as a two-body interaction with a scattering length of  order
the particle separation.

\subsection{Induced interactions}

In a multi-component system the effective interaction between atoms of one
species is altered by the presence of the other components.  As an example
of this we return to the stability condition (\ref{stab2}).
Since the boson gas must be
stable to density fluctuations when the fermions are uniform, the
compressibility of the bosons must be positive, which means that
$\partial \mu_B/\partial n_B >0$. Also, ${\partial \mu_B}/{\partial n_F}=
{\partial\mu_F}/{\partial n_B}={\partial^2 E}/{\partial n_B \partial n_F}$. 
Thus the stability condition  (\ref{stab2}) may be written as
\begin{equation}
\frac{\partial \mu_F}{\partial n_F} -\left(\frac{\partial \mu_F}{\partial
n_B}\right)^2\frac{\partial n_B}{\partial \mu_B}\ge 0.
\end{equation}
The first term here is the variation of the chemical potential of the
fermions
when the fermion density is changed, keeping the boson density fixed. The
second
term may be regarded as an induced interaction,
due to the fact that one fermion
tends to attract or repel bosons,
depending on the sign of the boson-fermion
interaction, and the change in the boson density changes the energy of
a second fermion.  The induced
interaction is analogous to the phonon-induced
attraction in metals.  Note that this interaction is always attractive,
irrespective of the sign of the boson-fermion interaction.  By use of the
chain rule, one may
show that the stability condition is equivalent to
\begin{equation}
\left.\frac{\partial \mu_F}{\partial n_F}\right|_{\mu_B} \ge 0.
\end{equation}
This way of looking at interactions has been exploited in the context of
liquid mixtures of $^3$He and $^4$He \cite{bbp}.
The induced interaction contribution to the stability criterion corresponds
to the diagram shown in Fig.\ 1, evaluated in the long-wavelength limit. 

\begin{figure}
\begin{center}
\epsfig{file=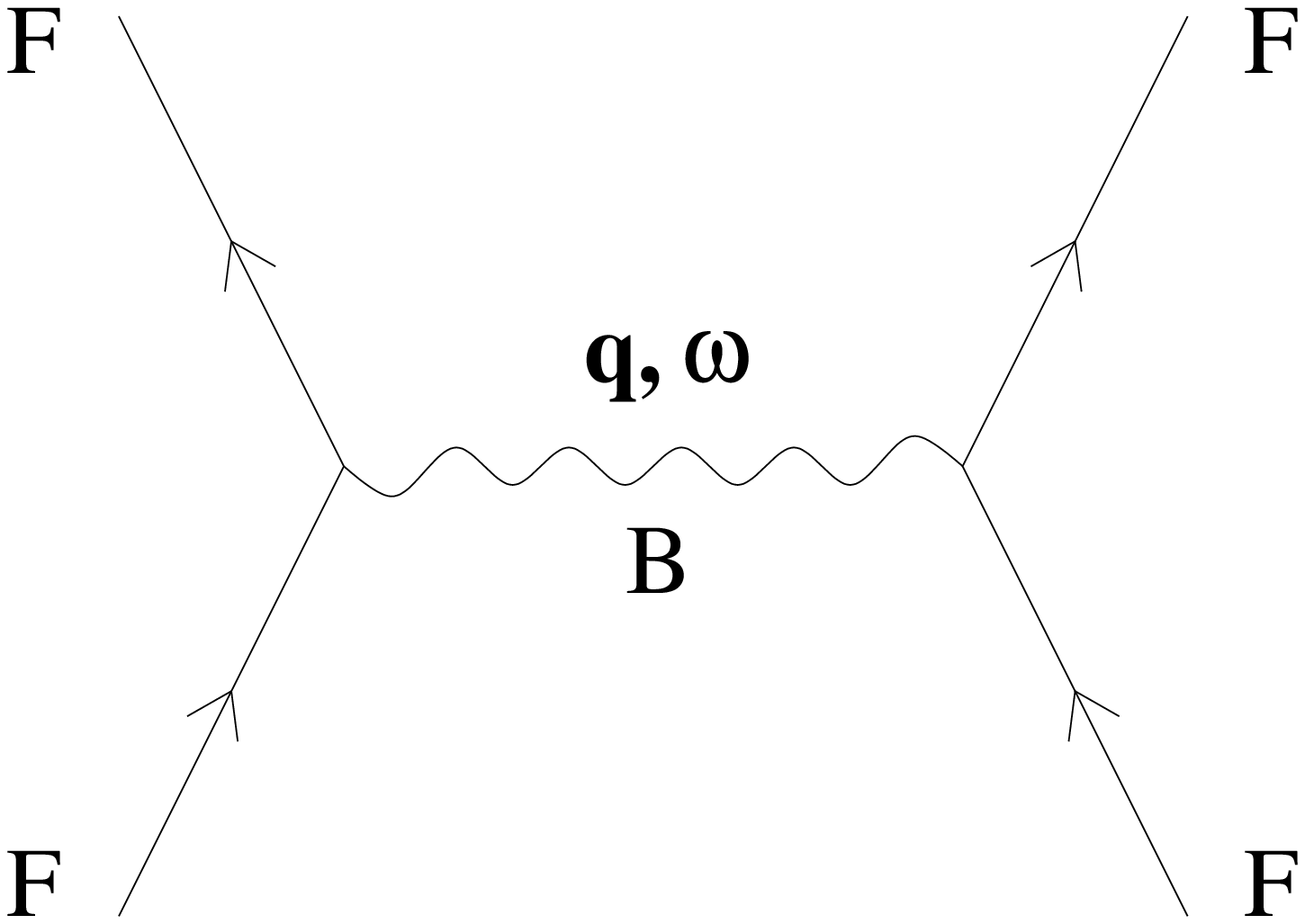,width=3.0cm,height=3.0cm,angle=-90}
\vspace{.2cm}
\begin{caption}
{Diagrammatic representation of the interaction between fermions induced
by exchange of a boson density fluctuation.}
\end{caption}

\end{center}

\label{?}
\end{figure}

For a dilute gas, the induced interaction is easily evaluated 
using the expression (\ref{enexpre}) for the energy, since
\begin{equation}
\frac{\partial \mu_F}{\partial n_B} = U_{BF}
\end{equation}
and
\begin{equation}
\frac{\partial \mu_B}{\partial n_B} = U_{BB}.
\end{equation}
Thus the induced interaction is given by
\begin{equation}
U_{\rm ind}= -\frac{U_{BF}^2}{U_{BB}},
\label{indint}
\end{equation}
and it is equivalent to an effective reduction of the
scattering length by an amount $(a_{BF}^2/a_{BB})(m_B m_F/m_{BF}^2)$.

At non-zero wavenumbers and frequencies the effective interaction is the
natural generalization of Eq.\ (\ref{indint}),
the static long-wavelength response
function $-\partial n_B/\partial \mu_B$ for the bosons being replaced by
the density-density response function $\chi_B(q, \omega)$.  Thus
\begin{equation}
 U_{\rm ind}(q, \omega)= {U_{BF}^2}\chi_B(q, \omega).
\end{equation}
In the Bogoliubov approximation the density-density
response function is given by
\begin{equation}
\chi(q, \omega) = \frac{n_B q^2}{m_B(\omega^2 -\omega_q^2)},
\end{equation}
where the excitation frequencies are the Bogoliubov ones
\begin{equation}
\hbar \omega_q =\left[\epsilon^0_q(\epsilon^0_q +2 n_B U_{BB})\right]^{1/2} ,
\end{equation}
with $\epsilon_q^0=\hbar^2q^2/2m_B$ being the free boson energy.
Thus the static induced interaction is
\begin{equation}
 U_{\rm ind}(q, \omega)= -U_{BF}^2 \frac{n_B}{n_B U_{BB} +
( \hbar^2q^2/4m_B)}.
\end{equation}
In coordinate space this is a Yukawa, or screened Coulomb,
interaction
\begin{equation}
U_{\rm ind}(r) = - \frac{m_B n_B U_{BF}^2}{\pi\hbar^2}
 \frac{e^{-\sqrt 2r/ \xi}}{r},
\end{equation}
where $\xi$ is the coherence (healing) length for the bosons, given by
\begin{equation}
\xi^2 = \frac{\hbar^2}{2 m_B n_B U_{BB}}.
\end{equation}

The induced interaction thus has the interesting feature
that at long wavelengths 
it gives rise to effects which are of the same 
order of magnitude as a typical 
bare interaction, if the boson-boson and boson-fermion 
interactions are of the 
same order of magnitude and boson and fermion 
masses are comparable. The 
reason for this is that even though the induced 
interaction involves two 
boson-fermion interactions, the density-density 
response function for the bosons 
at long wavelengths is inversely proportional to 
the boson-boson interaction. 
At wave numbers greater than the inverse of the coherence 
length for the bosons, 
the magnitude of the induced interaction is reduced, 
since the boson 
density-density response function has a magnitude
$\sim 2n_B/\epsilon_q^0$. The induced interaction is thus strongest 
for momentum transfers less 
than $m_B s_B$, where $s_B =(n_B U_{BB}/m_B)^{1/2}$ 
is the sound speed in the boson gas. 
If bosons and fermions have comparable masses and densities, 
and the scattering 
lengths are comparable, for momentum transfers 
of order the Fermi momentum the 
interaction will be of order $4\pi \hbar^2 k_Fa^2/(2m)$. 
The induced interaction is therefore one power of the 
``diluteness parameter" $k_F a$ less than a typical direct 
interaction. For mixtures of two species of fermions (for 
example two different hyperfine states) with bosons the 
induced interaction is attractive, and therefore it will 
increase the transition temperature to a BCS superfluid state. 
Detailed calculations are necessary to determine how important 
the effect is quantitatively.

\section{Phase equilibrium}
\label{gencom}

The stability considerations given in the previous section 
are valid only for small changes.
We now solve the general stability problem
by considering the  conditions under which mixtures
with different concentrations of bosons and fermions may  be
in equilibrium with each other.

We  consider a mixture of $N_B$ bosons and $N_F$
fermions in a box of volume
$V$. These components may
either mix uniformly or form distinct phases,
which we label by the index $i$.
 If we ignore interpenetration
effects, a possible phase-separated configuration may be described by the
number of phases present, $I$, the bosonic and fermionic densities in each
phase, $n_{B,i}$ and $n_{F,i}$, and the
fractions of the total volume they occupy,
$v_i$. Since the total number of
particles is given, the following relations must
hold: $\sum_{i=1}^I n_{F,i}v_i=n_F=N_F/V$,
$\sum_{i=1}^I n_{B,i}v_i=n_B=N_B/V$,
and $\sum_{i}^Iv_i=1$.
When $I=1$ we recover the case of a homogeneous
mixture. 
Let us now turn our attention to the case $I=2$.
The total energy is the sum of contributions due to boson-boson
interactions, boson-fermion interactions and the kinetic
energy of the fermions,
\begin{equation}
{\cal E}=\sum_{i=1}^2 {\cal E}_{i}=V\sum_{i=1}^2v_iE_i,
\label{e200}
\end{equation}
where
\begin{equation}
E_i= \frac{1}{2}n_{B,i}^2U_{BB}+
n_{B,i}n_{F,i}U_{BF}+\frac{3}{5}\epsilon_{F,i}n_{F,i},
\label{e201}
\end{equation}
with the Fermi energies $\epsilon_{F,i}$ being given by
\begin{equation}
\epsilon_{F,i}=(6\pi^2)^{2/3}n_{F,i}^{2/3}\frac{\hbar^2}{2m_F}.
\label{fermien}
\end{equation}

The pressure $p_i$ in each phase is then found by differentiating
the energy with respect to the volume $V_i=Vv_i$ occupied by each phase,
\begin{equation}
p_i=-\frac{\partial {\cal E}_i}{\partial V_i}=
\frac{1}{2}n_{B,i}^2U_{BB}+
n_{B,i}n_{F,i}U_{BF}+\frac{2}{5}\epsilon_{F,i}n_{F,i},
\label{pressures}
\end{equation}
using the volume dependence of the Fermi energy given by (\ref{fermien}).
The first requirement for equilibrium is that $p_1=p_2$.
The two other conditions involve
the chemical potentials which are given by
\begin{eqnarray}
\nonumber \mu_{B,i} &=& \frac{\partial E_i}{\partial n_{B,i}}
=n_{B,i}U_{BB}+n_{F,i}U_{BF},\\
\mu_{F,i} &=& \frac{\partial E_i}{\partial n_{F,i}}
=\epsilon_{F,i}+n_{B,i}U_{BF}.
\label{muf1000}
\end{eqnarray}
If the boson density $n_{B,i}$ is non-zero in both
phases, then the chemical potentials $\mu_{B,i}$
must be equal. If the boson density vanishes
in one phase, then
the boson chemical potential in that phase must be higher
than in the other one, in order for the system to be in equilibrium.
The same is of course  true for the fermions.

We remark that for a two-component system it is not possible to have
3 or more distinct phases in equilibrium. This is seen as follows. 
Consider a situation where there are $I$ distinct mixed phases. The 
total number of conditions to be met is $3(I-1)$. If one 
wishes equilibrium to be possible over a range of parameters one
may in addition fix another variable, such as the total pressure.
The total number of conditions to be satisfied is then $3I-2$. However
the number of independent variables is just $2I$, corresponding to the 
boson and fermion densities for the $I$ phases. It is then clear that 
solutions to the equations are only possible if $I\leq 2$.
Similar arguments apply if some pure phases are present.

We thus consider a system consisting of two phases with
densities $n_{B,1}, n_{F,1}$
and $n_{B,2}, n_{F,2}$, with the first one occupying
 a share $v$ of the total volume.
Due to the conditions on the chemical potentials one has to
distinguish four cases, which must be analyzed one by one:

\vspace{.1cm}

\noindent A. {\em Two pure phases}:
The bosons and fermions are completely separated corresponding to
 $n_{F,1}=0$,
$n_{B,2}=0$ and $n_{B,1}\neq 0$, $n_{F,2}\neq 0$.

\noindent B. {\em A mixed phase and a purely fermionic one}:
The boson density vanishes in one region
corresponding to  $n_{F,1}\neq 0$,
$n_{B,2}=0$ and $n_{B,1}\neq 0$, $n_{F,2}\neq 0$.

\noindent C. {\em A mixed phase and a purely bosonic one}:
The fermion density vanishes in one region
corresponding to the conditions $n_{F,1} = 0$,
$n_{B,2}\neq 0$ and $n_{B,1}\neq 0$, $n_{F,2}\neq 0$.

\noindent D. {\em Two mixed phases}: All
densities $n_{B,1}$, $n_{F,1}$
and $n_{B,2}$, $n_{F,2}$ are different from zero,
while $n_{B,1}\neq n_{B,2}$ and $n_{F,1}\neq n_{F,2}$.

\vspace{.1cm}

First we consider cases A and B
and identify the two regions of the
$n_B$-$n_F$ plane in which they can occur. We  demonstrate below
that these regions are in fact different
and that in each of them the phase-separated configuration
is energetically favored compared to the homogeneous one.
 Subsequently we prove that the cases C and D are not realizable.

\subsection{Two pure phases}
\label{2pure}

As mentioned previously the pressures in the two phases must be the same.
These are found from the general formula (\ref{pressures}) by setting
$n_{F,1}=0$ in the expression for $p_1$ and $n_{B,2}=0$
in the expression for $p_2$. The condition  $p_1=p_2$ then becomes
\begin{equation}
\frac{1}{2}n_{B,1}^{2}U_{BB}=\frac{2}{5}An_{F,2}^{5/3},
\label{nb12pure}
\end{equation}
which shows that there is only one allowed
val'ue for $n_{B,1}$ for a given fermion density $n_{F,2}$.

Let us now look at the
chemical potentials. If the chemical potential in the fermionic phase
2 has to be lower or at most equal to that in the bosonic
phase 1, then it follows from (\ref{muf1000}) that
\begin{equation}
\epsilon_{F,2}\leq n_{B,1}U_{BB}.
\label{aa}
\end{equation}
Likewise, if the chemical potential
in the bosonic phase has to be lower than in the
fermionic one, then
\begin{equation}
n_{B,1}U_{BB}\leq n_{F,2}U_{BF},
\label{ab}
\end{equation}
which shows that $U_{BF}$ must be positive.
Inserting the value of $n_{B,1}$ from (\ref{nb12pure})
into (\ref{aa}) and (\ref{ab})  we obtain from (\ref{aa}) the  condition
\begin{equation}
n_{F,2}\geq \left(\frac{5AU_{BB}}{4U_{BF}^2}\right)^3,
\label{nf2gt}
\end{equation}
while (\ref{ab}) yields
\begin{equation}
n_{F,2}\geq \left(\frac{4AU_{BB}}{5U_{BF}^2}\right)^3,
\end{equation}
which is less restrictive than (\ref{nf2gt}).
Hence equilibrium of a pure boson phase and a pure fermion one 
is possible if and only if the fermion density satisfies
(\ref{nf2gt})  and $n_{B,1}$ is given by
(\ref{nb12pure}).

Now, since
\begin{eqnarray}
\nonumber n_B &=& v\, n_{B,1}\;\;\; {\rm and}\\
n_F &=& (1-v)\, n_{F,2},  \label{totdens}
\end{eqnarray}
where $v$ is the fraction
of the total volume occupied by the bosons, we can find those
total densities for which the system can
completely separate by using the allowed
values of $n_{F,2}$ and
$n_{B,1}$ in (\ref{totdens}).
These densities correspond to the points above the
uppermost full line in the $n_B$ vs. $n_F$ plot in Fig.\ 3.

Let us now compare the energy
of a phase-separated  configuration with that of a mixed state
with the same number of particles, as given by (\ref{totdens}).
According to (\ref{enexpre}) the energy  of the mixed system is
\begin{equation}
{\cal E}_{\rm mix}=V\left(\frac{1}{2}n_B^2U_{BB}+n_Bn_FU_{BF}
+\frac{3}{5}n_F\epsilon_F\right),  \label{emix}
\end{equation}
where the densities $n_B$ and $n_F$ are given by (\ref{totdens}).
On the other hand, the energy  of the phase-separated system
with the same number of particles is
\begin{equation}
{\cal E}_{\rm sep}=V\left(\frac{1}{2}n_{B,1}^2vU_{BB}
+\frac{3}{5}\epsilon_{F,2}n_{F,2}(1-v)\right).
\end{equation}
The energy difference ${\cal E}_{\rm mix}-{\cal E}_{\rm sep}$
is obtained by substituting for $n_B$ and $n_F$ from (\ref{totdens}),
with $n_{B,1}$ given by (\ref{nb12pure}),
\begin{eqnarray}
\nonumber \frac{1}{V}({\cal E}_{\rm mix}-{\cal E}_{\rm sep})&=&
-\frac{2}{5}v(1-v)An_{F,2}^{5/3}\\
\nonumber &+&v(1-v)n_{F,2}^{11/6}U_{BF}(4A/5U_{BB})^{1/2}\\
&-&\frac{3}{5}(1-v)(1-(1-v)^{2/3})A  n_{F,2}^{5/3}.
\end{eqnarray}
 Since by (\ref{nf2gt}) we must have $n_{F,2}^{1/6}
\geq \sqrt{5AU_{BB}/4U_{BF}^2}$, then
\begin{equation}
\frac{1}{V}({\cal E}_{\rm mix}-{\cal E}_{\rm sep})\geq
 \frac{3}{5}A  n_{F,2}^{5/3}(1-v)^{5/3}(1-(1-v)^{1/3})\geq 0
\end{equation}
for any $v$ between 0 and 1,
and the equality holds only at $v=0$ and $v=1$.
So in the density ranges where equilibrium is possible ${\cal E}_{\rm sep}$
is always less than ${\cal E}_{\rm mix}$,
and phase separation is energetically favored compared
to mixing.

\subsection{A mixed phase and a purely fermionic one}
\label{1fer}

We now consider  a mixed phase
in equilibrium with a pure fermion phase and
let $n_{B,2}=0$. As we shall see in the following, this type
of phase separation is stable in a different range of
total densities than that for case A.
The pressure equilibrium condition takes the form
\begin{equation}
\frac{1}{2}n_{B,1}^2U_{BB}+
n_{B,1}n_{F,1}U_{BF}+\frac{2}{5}\epsilon_{F,1}n_{F,1}
= \frac{2}{5}\epsilon_{F,2}n_{F,2}.
\label{pre1fer}
\end{equation}
Beside this we have to impose also the equality of the fermion chemical
potentials in the two phases, which implies
\begin{equation}
n_{B,1}=\frac{A}{U_{BF}}(n_{F,2}^{2/3}-n_{F,1}^{2/3}).
\label{chf1fer}
\end{equation}
First let us assume that  $U_{BF}$ is positive.
Then  (\ref{chf1fer}) implies that
$n_{F,2}\geq n_{F,1}$,
 and it  allows us to eliminate $n_{B,1}$ from equation
(\ref{pre1fer}). The resulting
equation is conveniently expressed in terms of the
ratio between the fermion densities,
\begin{equation}
x=\left(\frac{n_{F,1} }{n_{F,2}}\right)^{1/3}
\label{x}
\end{equation}
and the dimensionless constant $\lambda$, defined by
\begin{equation}
\lambda=\frac{AU_{BB}}{U_{BF}^2n_{F,2}^{1/3}}.
\label{lambda}
\end{equation}
Inserting $n_{B,1}$ from (\ref{chf1fer}) into (\ref{pre1fer})
we thus obtain
\begin{equation}
\frac{\lambda}{2}(1-x^2)^2=\frac{2}{5}-x^3+\frac{3}{5}x^5.
\label{eq1fer}
\end{equation}
By solving this equation one obtains
 $n_{f,1}$ as a function of $n_{f,2}$.
The problem is greatly simplified by noticing that $(1-x)^2$ is a
common factor on both sides of Eq.\ (\ref{eq1fer}). The value $x=1$
implies that the fermion density is the same everywhere,
while the boson density vanishes everywhere according to eq. (\ref{chf1fer}). 
The solution $x=1$ can therefore be rejected,
and (\ref{eq1fer})  becomes the cubic equation
\begin{equation}
6x^3 +(12-5\lambda)x^2+(8-10\lambda)x+4-5\lambda=0
\label{cubic}
\end{equation}
Since we have assumed that $U_{BF}$ is positive or zero,
which, according to (\ref{chf1fer}), implies that $n_{F,1}\leq n_{F,2}$,
the physical range of $x$ lies between 0 and 1. Evidently
$\lambda=4/5$ yields the solution $x=0$, while $x=1$ is a solution
provided $\lambda=3/2$. The cubic equation (\ref{cubic}) has one
positive root in the range $0<x<1$, when $\lambda$ lies in the
range
\begin{equation}
\frac{4}{5}<\lambda<\frac{3}{2}.
\label{interval2}
\end{equation}
According to  (\ref{lambda}) this implies that
the fermion density $n_{F,2}$ is bounded from below as
well as from above. We conclude that
to each value of $n_{F,2}$ lying within the range
 (\ref{interval2}),
there corresponds  one value of $n_{F,1}$ and, from Eq.\ (\ref{chf1fer}),
one value of $n_{B,1}$.

In Fig.\ 2 we plot the values of $n_{F,1}$ and $n_{B,1}$ found for each
 $n_{F,2}$.

\begin{figure}
\begin{center}
\epsfig{file=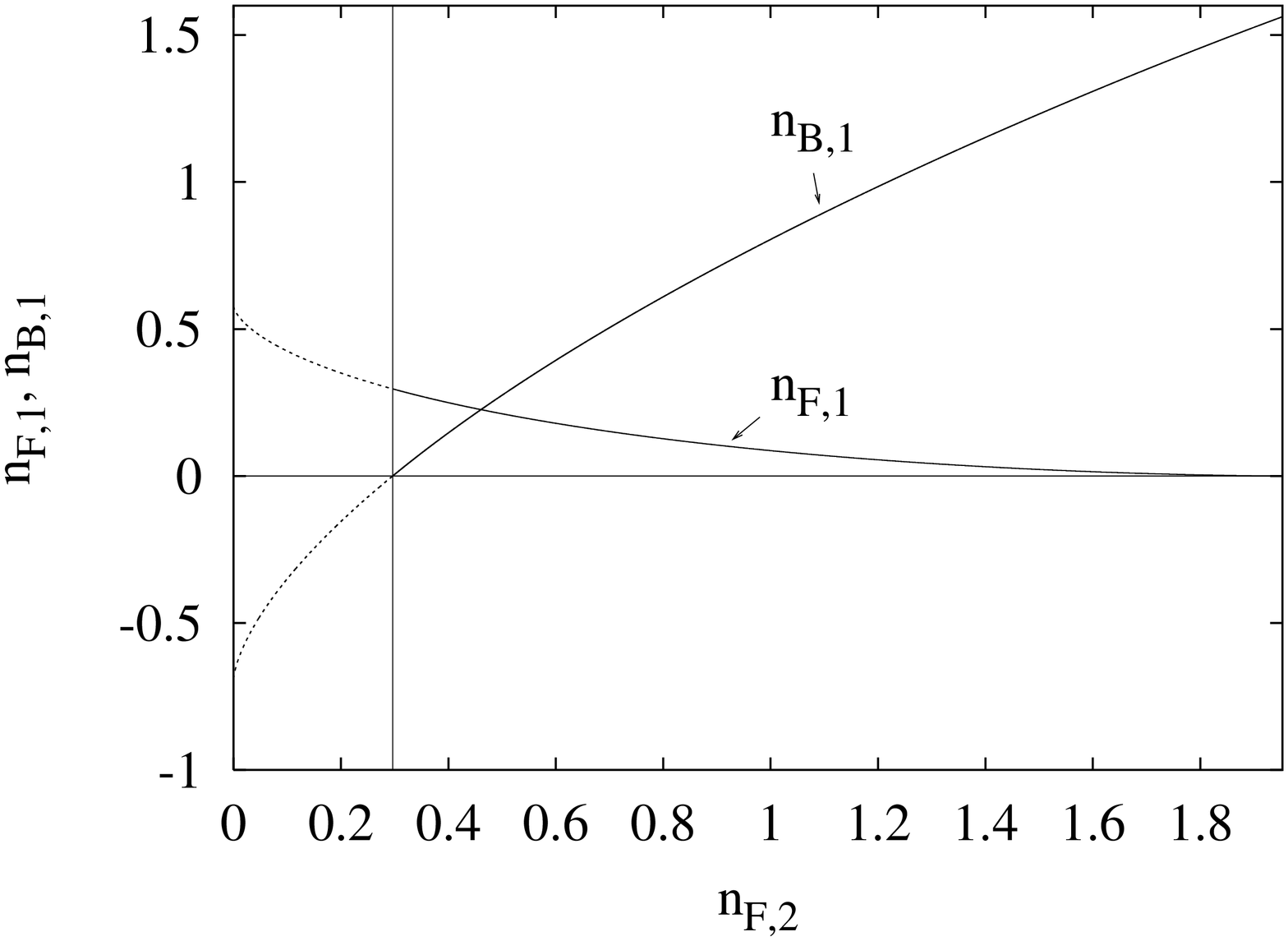,width=7.0cm,height=\linewidth,angle=-90}
\vspace{.2cm}
\begin{caption}
{Densities of fermions ($n_{F,1}$) and bosons ($n_{B,1}$) in
a mixed phase in equilibrium with a pure fermion phase of density $n_{F,2}$.
 The fermion densities are in units of $A^3U_{BB}^3/U_{BF}^6$, 
and the boson one in units of 
$A^3U_{BB}^2/U_{BF}^5$. The vertical line corresponds to the value
$(2/3)^3$, for which $n_{F,1}=n_{F,2}$ and $n_{B,1}=0$.
The dashed parts of the curves are unphysical for $U_{BF}>0$,
since the boson density is negative. For negative $U_{BF}$ they correspond 
to solutions with positive boson density, since the boson densities in the
figure are then scaled to a negative quantity. However for the mixed phase 
under these conditions $n_{F,1}>(2/3)^3A^3U_{BB}^3/U_{BF}^6$ 
and it is unstable.} 
\end{caption}

\end{center}

\label{???}
\end{figure}

When a mixture is separated into two phases with one allowed set of $n_{F,2}$,
$n_{F,1}$ and $n_{B,1}$, it is straightforward to show that the boson chemical
potential is always less in the mixed phase than in the purely fermionic one,
but for brevity we omit the proof.

Just as we did in the previous section,  we can deduce the
set of all total densities which could
undergo a phase separation of this type.
Since
\begin{eqnarray}
\nonumber n_B &=& v\, n_{B,1}\;\;\; {\rm and}\\
n_F &=& v\, n_{F,1}+(1-v)\, n_{F,2},  \label{nbnf5}
\end{eqnarray}
they are obtained by using allowed values for $n_{F,2}$, $n_{F,1}$ and
$n_{B,1}$,
and varying $v$ from zero to one.
These densities correspond to the points in the
region between the upper and lower full curves in Fig.\ 3.
The lower line is obtained
by letting $v=1$ and is therefore
just the set of allowed $n_{B,1}$ and $n_{F,1}$.
The upper one is found by setting $n_{F,2}=(5/4)^3(AU_{BB}/U_{BF}^2)^3$,
the maximum value according to (\ref{interval2}), 
to which correspond $n_{F,1}=0$ and
$n_{B,1}=(5/4)^2(A^3U_{BB}^2/U_{BF}^5)$, and varying $v$.
It should be noted that this upper curve
is exactly the one
above which complete separation into a pure boson phase
and a pure fermion one  could take place. This proves the
statement that cases A and B are
stable for different values of total densities,
which in view of the proof on the
energy of case A, also demonstrates that above
the upper line of Fig.\ 3 the system will be completely phase separated.

Also in the present case we can show that the phase-separated configuration
has lower energy than the corresponding mixed one. The
proof proceeds as in case A and we shall therefore omit it.

\begin{figure}
\begin{center}
\epsfig{file=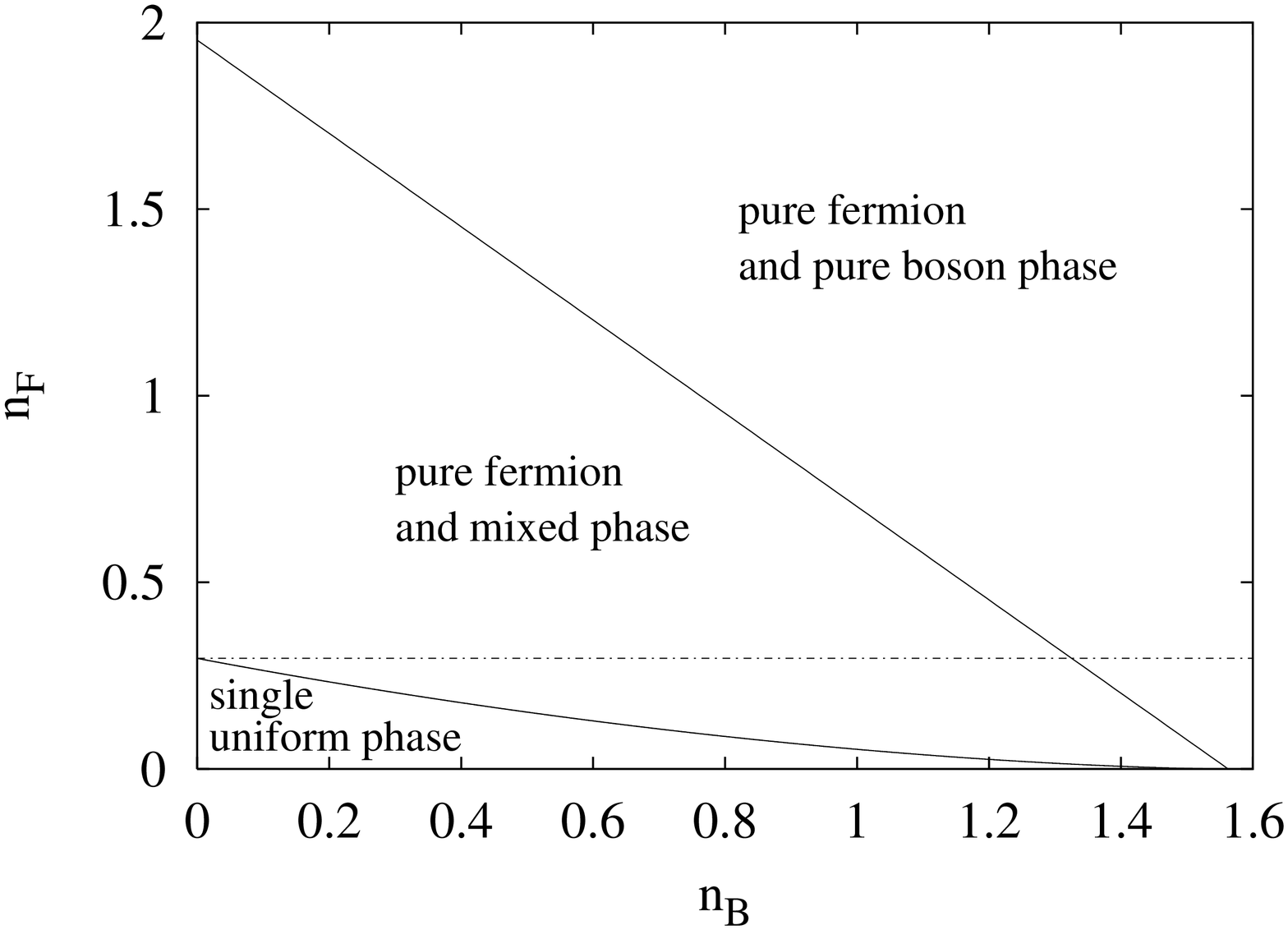,width=7.0cm,height=\linewidth,angle=-90}
\vspace{.2cm}
\begin{caption}
{Phase diagram of uniform boson-fermion mixtures. The total fermion density
$n_{F}$ is given in units of
$A^3U_{BB}^3/U_{BF}^6$, and the total boson density $n_B$ 
in units of $A^3U_{BB}^2/U_{BF}^5$.
The horizontal line corresponds to the value $n_F=(2/3)^3A^3U_{BB}^3/U_{BF}^6$, 
above which the uniform system is unstable to small fluctuations.}
\end{caption}

\end{center}

\label{??}
\end{figure}

One more comment is due at this point.
In the linear stability analysis we found
that the homogeneous system would become
unstable at densities violating (\ref{linsta}), but this
value of the fermion density is just where the lower
curve in Fig.\ 3 meets the vertical axis.

Finally let us consider the situation when $U_{BF}$ is negative.
In this case  the fermion density in phase 1 must be
greater than that in phase 2,
$n_{F,1}>n_{F,2}$, according to (\ref{chf1fer}). We find again
that the density ratio $x=n_{F,1}/n_{F,2}$ must satisfy (\ref{cubic}), but
now $x$ lies in the region between 1 and infinity.
By solving the cubic equation (\ref{cubic})
it is readily shown (compare Fig.\ 2) that the
values obtained for $n_{F,1}$
are now {\it greater} than that which is required to
ensure linear stability in phase 1. The phase-separated
configuration is thus  unstable for $U_{BF}<0$.
The  attractive boson-fermion interaction in
phase 1 leads to a collapse of the
configuration with a mixed and a purely fermionic phase,
presumably to a state where the bosons and
fermions are clumped together and where the total energy is no
longer given by (\ref{enexpre}).

With this our knowledge of the phase diagram is complete, since
the cases  C and D cannot be realized, as proven below.

\subsection{Two mixed phases}
\label{2mix}

We shall now demonstrate that the cases C and D cannot be realized.
First we consider  two phases with different non-zero boson and
fermion densities.

The condition for  equality of pressures is
\begin{eqnarray}
\nonumber \frac{1}{2}(n_{B,1}^2&-&n_{B,2}^2)U_{BB}+
(n_{B,1}n_{F,1}-n_{B,2}n_{F,2})U_{BF}\\
&+&\frac{2}{5}(\epsilon_{F,1}n_{F,1}-\epsilon_{F,2}n_{F,2})=0,
\label{pre2mix}
\end{eqnarray}
and for equality of the  chemical potentials
\begin{equation}
(n_{B,1}-n_{B,2})U_{BB}=(n_{F,2}
-n_{F,1})U_{BF},                   \label{mub2mix}
\end{equation}
for bosons and
\begin{equation}
(n_{B,1}-n_{B,2})U_{BF}=A(n_{F,2}^{2/3}-
n_{F,1}^{2/3}).    \label{muf2mix}
\end{equation}
 for  fermions. Combining equation (\ref{mub2mix}) with (\ref{muf2mix})
and using the definitions (\ref{x}) and (\ref{lambda}) we get
\begin{equation}
1-x^3=\lambda(1-x^2).           \label{1}
\end{equation}
Substituting (\ref{mub2mix})
into (\ref{pre2mix}) and rearranging  terms, we obtain
furthermore
\begin{equation}
(1-x^6)=\frac{4}{5}\lambda(1-x^{5}).  \label{2}
\end{equation}
By writing $(1-x^6)=(1+x^3)(1-x^3)$
and inserting $1-x^3$ from (\ref{1}) into (\ref{2}) we obtain
\begin{equation}
-\frac{1}{5}x^5+\frac{1}{5}+ x^3-x^2=0.
\label{fifth}
\end{equation}
The left hand side of (\ref{fifth})
has a triple root at $x=1$, and (\ref{fifth}) can
therefore  be written as
\begin{equation}
\frac{1}{5}(x^2+3x+1)(x-1)^3=0
\end{equation}
which clearly shows, since $x$ is positive, that the only solution is
the trivial one, $x=1$. We have thus proven that two mixed phases cannot be
in equilibrium with each other.

\subsection{A mixed phase and a purely bosonic one}
\label{1bos}

Finally we consider the case when one of the phases, say phase 2,
is free of fermions and prove that equilibrium is impossible.
We observe that  (\ref{pre2mix}) and (\ref{mub2mix}) are valid
also in this case, if we put $n_{F,1}=0$.
The same is true also for
equation (\ref{2}), derived by combining them, which
upon the use of $x=0$ becomes
\begin{equation}
\lambda=\frac{5}{4}.
\label{condlam}
\end{equation}
According to the definition (\ref{lambda}) of $\lambda$ the
system can therefore only be in equilibrium
for one particular value of the fermion density.
However, the  fermion chemical
potential in region 1 must be greater than that in region 2,
\begin{equation}
 n_{B,1}U_{BF}\geq An_{F,2}^{2/3} +n_{B,2}U_{BF}.
\end{equation}
Substituting for $n_{B,1}-n_{B,2}$, as derived from equation (\ref{mub2mix})
with $n_{F,1}=0$, we obtain
\begin{equation}
n_{F,2}\frac{U_{BF}^2}{U_{BB}} \geq An_{F,2}^{2/3},
\end{equation}
which according to the definition (\ref{lambda})
implies  $\lambda\leq 1$. But the solution (\ref{condlam})
does not satisfy this condition and must therefore be discarded.
We conclude that case D also cannot be realized physically.

\section{Discussion and conclusions}
\label{discussion}

As we have emphasized, the properties of
boson-fermion mixtures are of interest
because such mixtures allow sympathetic cooling of fermions, and because the
boson-induced fermion-fermion interaction,
being attractive, favors pairing of
fermions, thus raising the BCS transition temperature.
In this concluding section we discuss the implications of
our  results  for actual experimental conditions.

Our main findings  are summarized in Fig.\ 3. This phase
diagram shows that fermions and bosons under most
circumstances will  form a mixed phase,
 unless the interaction parameters assume quite large values.
We shall now  quantify this statement, using lithium,
potassium and rubidium atoms as concrete examples.
From the results obtained above,  for instance the stability condition
(\ref{nf2gt}),
it is apparent that the value of the dimensionless
fermion density
$n_F (U_{BF}^2/AU_{BB})^3 $ plays a key role
in determining whether  the fermions and bosons
mix or phase-separate into either a purely fermionic phase together
with a mixed phase or a purely fermionic together with a purely
bosonic phase (cf.\ Fig.\ 3).
The corresponding dimensionless value of the boson density
is according to (\ref{nb12pure}) given by $n_B (U_{BF}^5/A^3U_{BB}^2)$.

Let us  therefore look
at the values of the actual densities for which these
dimensionless densities  are equal to one.
A mixture of $^7$Li (boson) and $^6$Li (fermion)
has  recently been produced \cite{florian} with $^7$Li
trapped in the hyperfine state
$\left. |1,-1 \right>$ which, contrary to the $\left. |2,2 \right>$
state used in \cite{bec3}, has
a positive scattering length $\approx 10$ a.u., see \cite{verhaar2}.
For a $^7$Li-$^6$Li collision, on the other hand, the scattering length is
$\approx 38$ a.u.\cite{verhaar}. From this we deduce that
$\tilde{n}_F=1$
corresponds to the density $n_F=9.3\cdot 10^{23}\,
{\rm m}^{-3}$,
which is several orders of magnitude higher than the
densities which
can presently be realized experimentally. When expressed in terms of
the boson-boson scattering length $a_{BB}$ the density is seen to be
$n_F=1.2\cdot 10^{-4}a_{BB}^{-3}$, indicating the
smallness of the gas parameter $n_{F}a_{BB}^3$.

Other interesting mixtures involve the potassium isotopes $^{39}$K,
$^{40}$K and $^{41}$K, of which the first and the last
are bosons. In these cases using the values for the scattering 
lengths given in \cite{cote}, 
we predict that phase separation effects would take place for
densities exceeding $a_{BB}^{-3}$, which not only are much higher than can be
 experimentally realized, but also fall outside
the range of applicability of the dilute
gas approximation.

A possible candidate for phase separation does, however, exist
 among the rubidium isotopes.
From the scattering lengths given in \cite{burke} we find
 that for $^{87}$Rb and $^{84}$Rb mixtures
the phase separation could take place at reasonable densities.
Using the estimates $a_{BB}\approx 100$ a.u.\ and $a_{BF} \approx 550$ a.u.,
we have $\tilde{n}_F=1$
 when $n_F= 8.1\cdot 10^{-6} a_{BB}^{-3}= 7.1\cdot 10^{19} \,
 {\rm m}^{-3}$, which is of the same order as the densities
 already realized in experiments with Rb isotopes.

It is also illuminating to
compare our findings for the uniform case
with previous calculations \cite{klaus,stringari} for
mixtures in a trap. Due to  the Fermi pressure,
the fermion density in a trap is usually much
smaller than the boson density, and it is therefore a
good approximation to assume that the bosonic cloud is unaffected
by the fermions. The fermions on the other hand see an effective potential,
which may be attractive or repulsive inside the boson cloud,
depending on the  potential parameters and the ratio $U_{BF}/U_{BB}$
of the interaction parameters \cite{klaus,stringari}. If for simplicity
the two trapping potentials are taken to be isotropic
with same force constants, the effective potential
is repulsive if $U_{BF}>U_{BB}$. In this case the
fermions are therefore expelled and form a shell outside the
boson condensate. If the effective potential is attractive,
corresponding to the condition $U_{BF}<U_{BB}$,
the fermions will reside inside the boson cloud.
From the discussion above it is clear that in most cases
the fermions inside the cloud will be fully
mixed with the bosons. If the interaction parameters are appropriately
chosen one then enters the phase-separated regime of our phase diagram, 
as is the case in the example shown in Fig.\ 2 of \cite{klaus}, where
a pure fermion phase forms at the center of the trap.
The phase diagram which we have  obtained here for the uniform case
 may thus serve as a useful guide
for understanding the general behaviour of trapped mixtures.

\end{document}